\documentclass[conference]{IEEEtran}
\usepackage{ifpdf}

\usepackage{cite}
\usepackage[cmex10]{amsmath}
\usepackage{amssymb}
\usepackage{graphicx,amsmath,amssymb}

\usepackage{algorithmicx}
\usepackage{algpseudocode}
\usepackage[tight,footnotesize]{subfigure}
\usepackage{url}

\let\pa\partial
\newcommand{\T}{\mathcal{T}}
\let\r\rho
\newcommand{\pai}{\operatorname{PartialIntegrate}}
\newcommand{\coef}{\operatorname{Coefficient}}
\newcommand{\expo}{\operatorname{HighestExponent}}

\begin{document}
%
\title{Symbolic Derivation of Mean-Field PDEs from Lattice-Based Models}

\author{\IEEEauthorblockN{Christoph Koutschan, Helene Ranetbauer, Georg Regensburger, Marie-Therese Wolfram}
\IEEEauthorblockA{Johann Radon Institute for Computational and Applied Mathematics (RICAM)\\
Austrian Academy of Sciences (\"OAW)\\
Altenberger Stra\ss e 69, 4040 Linz, Austria\\
E-mail: christoph.koutschan@ricam.oeaw.ac.at,
helene.ranetbauer@oeaw.ac.at,\\
georg.regensburger@oeaw.ac.at,
mt.wolfram@ricam.oeaw.ac.at}}
 

\maketitle

\begin{abstract}

Transportation processes, which play a prominent role in the life and social
sciences, are typically described by discrete models on lattices.  For
studying their dynamics a continuous formulation of the problem via partial
differential equations (PDE) is employed.  In this paper we propose a symbolic
computation approach to derive mean-field PDEs from a lattice-based model.  We
start with the microscopic equations, which state the probability to find a
particle at a given lattice site. Then the PDEs are formally derived by Taylor
expansions of the probability densities and by passing to an appropriate limit
as the time steps and the distances between lattice sites tend to zero. We
present an implementation in a computer algebra system that performs this
transition for a general class of models.  In order to rewrite the mean-field
PDEs in a conservative formulation, we adapt and implement symbolic
integration methods that can handle unspecified functions in several
variables.  To illustrate our approach, we consider an application in crowd
motion analysis where the dynamics of bidirectional flows are studied.
However, the presented approach can be applied to various transportation
processes of multiple species with variable size in any dimension, for
example, to confirm several proposed mean-field models for cell motility.
\makeatletter
\def\blfootnote{\gdef\@thefnmark{}\@footnotetext}
\makeatother
\blfootnote{\copyright\ 2015 IEEE. Personal use of this material is
  permitted. Permission from IEEE must be obtained for all other uses, in any
  current or future media, including reprinting/republishing this material for
  advertising or promotional purposes, creating new collective works, for
  resale or redistribution to servers or lists, or reuse of any copyrighted
  component of this work in other works.
  This is the author's accepted version; the final publication is available at
  http://ieeexplore.ieee.org, DOI: 10.1109/SYNASC.2015.14}
\end{abstract}


%
\IEEEpeerreviewmaketitle

\section{Introduction}

Mean-field models play an important role in applied mathematics and have
become a popular tool to describe transportation dynamics in the life and
social sciences. In the derivation of such models the effect of a large number
of individuals on a single individual is approximated by a single averaging
effect, the so called mean-field. Applications include cell migration at high
densities, cf.~\cite{FLS2010,SBM2011}, transport across cell membranes as
occurring in ion channels, cf.~\cite{BDPS2010,BSW2012}, traffic
flow~\cite{LHR1955} as well as the motion of large pedestrian crowds, see
e.g.~\cite{BMP2011, BDMW2014}. Understanding the complex dynamics of large
interacting groups of particles is of high practical relevance and initiated a
lot of research in the field of physics, transportation research, and applied
mathematics.

Mathematical models on the micro- as well as the macroscopic level have been
used successfully to describe various aspects of these transportation
processes.  On the microscopic level the dynamics of each individual are
modelled taking into account its interactions with all others as well as
interactions with the physical surrounding. This approach results in
high-dimensional and very complex systems of equations.  On the macroscopic
level the crowd is treated as a density which evolves according to a partial
differential equation (PDE) or systems thereof. The transition from the
microscopic to the corresponding macroscopic description is an active area of
research with a lot of open analytic questions.

On the microscopic level a distinction is made between two different models:
\textit{force-based} or \textit{lattice-based} models. In the former the
dynamics of each individual is determined by the forces acting upon it,
i.e. exerted from the others and the surrounding; the latter states the
probability to find a particle at a discrete position in space (the lattice
point) given the transition rates of the particle to move from one discrete
lattice point to another. 

Lattice-based models, also known as cellular automata, are a very prominent
tool to describe cell motility, cf.~\cite{SMLH2007}, as well as pedestrian
dynamics (cf.~\cite{blue2001cellular},~\cite{burstedde2001simulation}), since
exclusion processes can be included naturally. In exclusion-based processes
each lattice site can be occupied by at most one individual, giving a simple
way to account for the finite particle size. In the last years there has been
an increasing interest in the derivation of the corresponding continuum
equations in both fields, see for example~\cite{SBM2011, FLS2010} in case of
cell dynamics or~\cite{BMP2011, BDMW2014} describing pedestrian dynamics. The
general structure of the resulting mean-field equations depends on the
transition rates, but common features include
\begin{enumerate}
\item their conservative nature; i.e. they are based on the assumption that
  the total mass is conserved;
\item an underlying gradient flow or perturbed gradient flow structure with
  respect to a certain metric; solutions of the first one correspond to minimizers of an
  energy functional with respect to a certain metric.
\end{enumerate}
These structural features allow to write the mean-field PDEs in terms of the
diffusivity and energy functionals; quantities which are of interest for
analysists. Therefore it is desirable to derive the mean-field equations in
this conservative form, although the general formulation is not unique.

Different strategies have been used to pass to the macroscopic limit, i.e. to
derive the corresponding continuum equations as the number of particles tends
to infinity, in either approach. In force-based models the macroscopic limit
can be derived using the so called BBGKY hierarchies, see for
example~\cite{C2000}, initially developed in the field of statistical
physics. In the case of stochastic underlying dynamics the derivation of the
mean-field description has been studied rigorously for simpler models by considering the
hydrodynamic limit, see \cite{KL2013}. Simpler models, such as the Patlak-Keller-Segel model for chemotaxis or reaction-diffusion
equations, were rigorously derived for a stochastic many-particle system, see \cite{S2000} and
\cite{O1989} respectively. \\
We would like to mention a related work by Penington and co-workers \cite{PHL2011}
on a systematic construction method to determine the continuum limit of nonlinear PDEs
from discrete lattice based models. Their approach is based on representing
the transition rates using appropriate rotation operators as well as symmetry
conditions to derive general expressions for the transportation coefficients in the
corresponding nonlinear PDEs. This technique can be used for a large class of 
problems (including multi-species dynamics in various space dimension), but assumes that
the transition rate of a species depends only on the average occupancy of a site by
any of the different species and not if the site is occupied by a particular subpopulation
or not. This approach cannot be applied to the pedestrian model presented later on, since the transition
rates depend on the affiliation to either group.

In the case of a lattice-based model we
\begin{enumerate}
\item replace the probability to find a particle at a lattice site by a formal
  Taylor expansion (up to a certain order) of the corresponding density,
\item pass to an appropriate limit as the lattice size and time tends to zero
  (dropping higher-order terms).
\end{enumerate}

In this paper we present an algorithmic approach to derive the corresponding
continuum equations from a lattice-based model using tools from symbolic
computation. While it is relatively straightforward to perform the formal
Taylor expansions and the corresponding limit, it is a more challenging task
to rewrite the PDEs obtained this way in a conservative form.  For this
purpose, we employ symbolic integration methods that can deal with unspecified
functions in several variables. Our approach allows us to deduce the
mean-field equations for a general class of transportation processes in
multiple space dimension, including the dynamics of multiple species that may
have different size or shape.  We illustrate our approach for a minimal model
of pedestrian dynamics, which includes cohesion and aversion in bidirectional
pedestrian flows.\\

We have produced a prototype implementation in Mathematica of the methods
described in this paper, which is available at
\url{http://www.koutschan.de/data/meanfield/}
together with a demo notebook.

\section{Formal derivation of a mean-field PDE model for bidirectional pedestrian dynamics}
\label{sec:derive}

We start with a specific example to illustrate the derivation of a mean-field
model from a discrete lattice-based approach in the case of bidirectional
pedestrian flows. We consider two groups of individuals --- one moving to the
right, the other to the left. The dynamics of each individual are determined
by cohesion and aversion --- this means they try to follow and stay close to
individuals moving in the same direction and to step aside when being
approached by an individual moving in the opposite direction. We expect that
these minimal dynamics will lead to the formation of directional lanes, a
phenomenon that has been observed in crowded corridors, pedestrian walks or
experiments.

\subsection{The microscopic model}

We start with the underlying microscopic model, i.e. a lattice-based approach
in which we consider, for the sake of simplicity, a rectangle $\Omega
\subseteq \mathbb{R}^2$ such as a corridor, partitioned into a square lattice
of grid size~$h$. This can be generalized to higher dimensions. Each lattice
site $(x_i, y_j) = (ih, jh)$, $i=0,\ldots, N$ and $j = 0, \ldots, M$ can be
occupied by an individual. We consider two groups moving in opposite direction
--- one to the right (called the reds) and one to the left (called the
blues). The probability to find a red individual at time $t$ at location
$(x_i,y_j)$ is given by:
\[
  r_{i,j}(t)= P(\text{red individual is at position } (x_i,y_j) \text{ at time t}),
\]
where $P$ denotes the probability. The probability for the blue individuals is
defined analogously. We denote by $\T_c^{\{i,j\}\rightarrow\{k,l\}}$ the
rate at which an individual of color~$c$ moves from $(x_i,y_j)$
to $(x_k,y_l)$.  The transition rates for the red and blue individuals
respectively are given by:
\begin{subequations}
\label{transprob}
\begin{align}\label{transprob1}
\begin{aligned}
\T_r^{\{i,j\}\rightarrow \{i+1,j\}}&=(1-\rho_{i+1,j})(1+\alpha \, r_{i+2,j}),\\
\T_r^{\{i,j\}\rightarrow \{i,j-1\}}&=(1-\rho_{i,j-1})(\gamma_0+\gamma_1 \, b_{i+1,j}),\\
\T_r^{\{i,j\}\rightarrow \{i,j+1\}}&=(1-\rho_{i,j+1})(\gamma_0+\gamma_2 \, b_{i+1,j}),
\end{aligned}
\end{align}
\begin{align}\label{transprob2}
\begin{aligned}
\T_b^{\{i,j\}\rightarrow \{i-1,j\}}&=(1-\r_{i-1,j})(1+\alpha \, b_{i-2,j}),\\
\T_b^{\{i,j\}\rightarrow \{i,j+1\}}&=(1-\r_{i,j+1})(\gamma_0 + \gamma_1 \, r_{i-1,j}),\\
\T_b^{\{i,j\}\rightarrow \{i,j-1\}}&=(1-\r_{i,j-1})(\gamma_0+\gamma_2 \, r_{i-1,j}),
\end{aligned}
\end{align}
\end{subequations}
where we write $\rho_{i,j}=r_{i,j}+b_{i,j}$, and with $0\leq \gamma_0,\gamma_1,\gamma_2\leq 1$, $0\leq\alpha\leq\frac{1}{2}$.  The prefactor
$(1-\rho)$ in all terms of \eqref{transprob} corresponds to the so-called size
exclusion, i.e. an individual cannot jump into the neighboring cell if it is
occupied.  We assume that the transition rates only depend on lattice sites in
direction of movement, a reasonable assumption when modeling the movement of
pedestrians.  The second factors in~\eqref{transprob} correspond to cohesion
and aversion. Cohesion is modelled in the first line in each case, by
introducing a factor $\alpha > 0$ which increases the probability to move in
the walking direction if the individual in front, i.e. in~\eqref{transprob1} at
position $(x_{i+2},y_{j})$, is moving in the same direction. The second and
third line in each case account for aversion via sidestepping. If an
individual, i.e. in~\eqref{transprob1} a blue particle located at
$(x_{i+1},y_j)$, is approaching, the red particle jumps up or down (with rates
$\gamma_1$ resp. $\gamma_2$). If $\gamma_1>\gamma_2$, the preference is to
jump to the right with respect to the direction of movement, if $\gamma_1<\gamma_2$, to the left. The parameter $\gamma_0>0$ corresponds to diffusion in the $y$-direction.

Then the evolution of the red particles is given by the so-called
master equation
\refstepcounter{equation}\label{master_eqns}
\begin{align}
&r_{i,j}(t_{k+1})=r_{i,j}(t_k)+\T_r^{\{i-1,j\}\rightarrow \{i,j\}} r_{i-1,j}\nonumber \\
&~~ + \T_r^{\{i,j+1\}\rightarrow \{i,j\}} r_{i,j+1}+\T_r^{\{i,j-1\}\rightarrow \{i,j\}} r_{i,j-1}
  \tag{\theequation a}\label{master_r} \\
&~~ -\bigl(\T_r^{\{i,j\}\rightarrow \{i+1,j\}}+\T_r^{\{i,j\}\rightarrow \{i,j-1\}}+\T_r^{\{i,j\}\rightarrow \{i,j+1\}}\bigr) r_{i,j}. \nonumber
\end{align}
Hence the probability to find a red particle at location $(x_i, y_j)$
corresponds to the probability that a particle located at $(x_{i-1},y_j)$
jumps forwards (first term), particles located above or below, i.e. at
$(x_i,y_{j \pm 1})$ jump up or down (second line), minus the probability that
a particle located at $(x_i,y_j)$ moves forward or steps aside (third
line). The evolution of the blue particles can be formulated analogously:
\begin{align}
&b_{i,j}(t_{k+1})=b_{i,j}(t_k)+\T_b^{\{i+1,j\}\rightarrow \{i,j\}} b_{i+1,j}\nonumber\\
&~~+ \T_b^{\{i,j-1\}\rightarrow \{i,j\}} b_{i,j-1} + \T_b^{\{i,j+1\}\rightarrow \{i,j\}} b_{i,j+1}
  \tag{\theequation b}\label{master_b} \\
&~~- \bigl(\T_b^{\{i,j\}\rightarrow \{i-1,j\}}+\T_b^{\{i,j\}\rightarrow \{i,j+1\}}+\T_b^{\{i,j\}\rightarrow \{i,j-1\}}\bigr) b_{i,j}. \nonumber
\end{align}

\subsection{Derivation of the macroscopic model}

In the next step we formally derive the limiting mean-field equations as the
grid size $h$ and the time steps $\Delta t$ tend to zero.  Hence we consider
the formal hyperbolic limit as $h=\Delta t=\Delta x=\Delta y \rightarrow 0$ in
Equations~\eqref{master_eqns}. We first substitute the transition
rates~\eqref{transprob} in Equation~\eqref{master_r} and obtain
\begin{align}\label{master_r1}
  r_{i,j}(t_{k+1}) - &r_{i,j}(t_k) =(1-b_{i,j}-r_{i,j}) (1+\alpha r_{i+1,j})r_{i-1,j} \nonumber \\
  +&(\gamma_0+\gamma_1 b_{i+1,j+1})(1-b_{i,j}-r_{i,j}) r_{i,j+1} \nonumber \\
  +&(\gamma_0+\gamma_2 b_{i+1,j-1}) (1-b_{i,j}-r_{i,j}) r_{i,j-1}\nonumber \\
  -&\bigl((1-b_{i+1,j}-r_{i+1,j})(1+\alpha r_{i+2,j}) \\
  +&(\gamma_0+\gamma_1 b_{i+1,j}) (1-b_{i,j-1}-r_{i,j-1})\nonumber \\
  +&(\gamma_0+\gamma_2 b_{i+1,j}) (1-b_{i,j+1}-r_{i,j+1})\nonumber
  \bigr)r_{i,j},
\end{align}
and a similar equation for the evolution of the blue particles. Next we employ
Taylor expansions up to second order of all the occurring probabilities. For
example, the probability to find a red particle at location $(x_{i+1},y_j)$
can be expanded as
\begin{align}
\begin{aligned}
  \label{tayl}
  r_{i+ 1,j} &= r_{i,j} + h\pa_xr_{i,j} + \tfrac12 h^2 \pa_x^2r_{i,j}  +\mathcal{O}(h^3).\\
\end{aligned}
\end{align}
After expanding all probability densities we keep the terms up to second order
and consider the formal limit as $\Delta t = \Delta x=\Delta y \rightarrow
0$. This leads to the following system of PDEs for the densities of the red and
blue particles, which can be either obtained by tedious hand calculations, or
by the computer-algebra methods described in Section~\ref{sec:symb}:
\begin{subequations}
\begin{align}
\pa_tr&=-\pa_x\left((1-\r)(1+\alpha r)r\right)+(\gamma_1-\gamma_2)\pa_y\left((1-\rho)b r\right)\nonumber\\
& \quad -\frac{h}{2}\left[\pa_x^2(r(1-\r)(1+\alpha r))-2\pa_x((1-\r)\pa_x r)\right]\nonumber\\
&\quad +\frac{h}2\left[(\gamma_1+\gamma_2)\pa_y\left((1-\r)\pa_y(rb)+br\pa_y\r\right)\right.\label{e:pdered}\\
&\qquad \quad +2\gamma_0\pa_y\left((1-\rho)\pa_y r +r\pa_y\rho\right)\nonumber\\
&\qquad \quad +\left. 2(\gamma_1-\gamma_2)\pa_y\left((1-\r)r\pa_xb\right)  \right],\nonumber\\[1em]
\pa_tb &=\pa_x\left((1-\r)(1+\alpha b)b\right)-(\gamma_1-\gamma_2)\pa_y\left((1-\rho)b r\right)\nonumber\\
& \quad -\frac{h}{2}\left[\pa_x^2(b(1-\r)(1+\alpha b))-2\pa_x((1-\r)\pa_x b)\right]\nonumber\\
&\quad +\frac{h}2\left[(\gamma_1+\gamma_2)\pa_y\left((1-\r)\pa_y(rb)+br\pa_y\r\right) \right.\label{e:pdeblue}\\
&\qquad \quad +2\gamma_0\pa_y\left((1-\rho)\pa_y b +b\pa_y\rho\right)\nonumber\\
&\qquad \quad \left.+2(\gamma_1-\gamma_2)\pa_y\left((1-\r)b\pa_x r\right)  \right].\nonumber
\end{align}
\label{e:pdesystem}%
\end{subequations}
The first terms on the right-hand side of~\eqref{e:pdered}
and~\eqref{e:pdeblue} result from the first-order terms in the Taylor
expansion. They correspond to the movement of the reds and blues to the right
and left respectively as well as the preference of either stepping to the
right or left (depending on the difference $\gamma_1 - \gamma_2$). The second
line corresponds to the second-order terms in x-direction, the last three
lines to the second-order terms due to side-stepping.  Note that Equations
\eqref{e:pdered} and~\eqref{e:pdeblue} can be written in a conservative form,
i.e. $\pa_t r =-\nabla \cdot F_r$ and $\pa_t b=-\nabla \cdot F_b$ for some
matrices $F_r$ and $F_b$. These so-called continuity equations are always
useful as they describe the transport of a conserved quantity, in our case
mass conservation.

\section{Algorithmic derivation of the\\ mean-field PDEs}
\label{sec:symb}

Symbolic computation, the field of mathematics that is concerned with
computer-implemented exact manipulation of mathematical expressions involving
variables/symbols, is meanwhile a well-established area of research and has
numerous applications.  Unfortunately, it is not as widely known as it should
be.  One reason may be that some applications are not straightforward and
require at least some insight or programming skills. But to those who get
moderately familiar with symbolic computation software, it becomes an
indispensable tool. There are plenty of general-purpose computer algebra
systems available, the most well-known being probably Mathematica, Maple, and
Sage. For our implementation we have chosen Mathematica.

In this section we demonstrate how the transition from the discrete
lattice-based model to a macroscopic PDE-based formulation is achieved using
techniques from symbolic computation.

\subsection{Expansion}

Recall that the lattice sites are given by $(x_i,y_j)=(ih,jh)$ for
$i,j\in\mathbb{Z}$; in the limit $h\to0$ one obtains the problem formulation
for the macroscopic model.  Let $r=r(x,y)$ and $b=b(x,y)$ denote the
densities of red and blue particles in the macroscopic model.  In order to
perform the transition from partial difference equations for $r_{i,j}$ and
$b_{i,j}$ to partial differential equations for $r(x,y)$ and $b(x,y)$, we
employ formal Taylor expansions of the probabilities appearing
in~\eqref{master_r1}, as discussed in Section~\ref{sec:derive}, for example:
\begin{subequations}
\begin{align}
  \label{tayl_x}
  r_{i+1,j} &= r + h\pa_xr + \tfrac12 h^2 \pa_x^2r + \dots = \sum_{k=0}^\infty \frac{h^k}{k!}\pa_x^kr, \\
  \label{tayl_y}
  b_{i,j+1} &= b + h\pa_yb + \tfrac12 h^2 \pa_y^2b + \dots = \sum_{k=0}^\infty \frac{h^k}{k!}\pa_y^kb.
\end{align}
\label{tayl_exps}%
\end{subequations}
Note that these calculations are done on a completely formal level.

Although the expansions~\eqref{tayl_exps} are not available as a built-in
command in Mathematica, it is a relatively simple task to implement them. We
have made some effort to design our implementation as general as
possible. This means that we do not fix the number of expansion variables
(this corresponds to the dimension of the domain~$\Omega$). Moreover, we allow
for discrete steps of any size, i.e, terms of the form $r_{i+a,j+b}$ with
$a,b\in\mathbb{Z}$ can be handled as well.

For our purposes it suffices to perform the Taylor expansions~\eqref{tayl_exps} on
the master equation~\eqref{master_r1} up to second order. While this is a
tedious calculation when done by hand, it is a trivial task for a computer
algebra system. Still, when writing the result in expanded form, we obtain a
huge expression for the right-hand side of~\eqref{master_r1}:
\begin{equation}\label{expand}
\begin{aligned}
  & r\pa_xb + \alpha r^2\pa_xb - \pa_xr + b\pa_xr + 2r\pa_xr + {}\\
  & \qquad\qquad
    + \langle 167\ \text{terms} \rangle - \tfrac18\gamma_2h^5r(\pa_y^2r)(\pa_x^2\pa_y^2b).
\end{aligned}
\end{equation}
Since $h$ is considered to be very small, all terms involving $h^2$ or
higher powers of~$h$ will be omitted (this corresponds to the polynomial
reduction modulo $h^2$).
In our example Mathematica returns the following expression:
\begin{equation}\label{polred}
\begin{aligned}
  & r\pa_xb + \alpha r^2\pa_xb - \pa_xr + b\pa_xr + 2r\pa_xr + {}\\
  & \qquad\qquad\qquad\qquad
    + \bigl\langle 56\ \text{terms} \bigr\rangle - \tfrac12\gamma_2hb^2\pa_y^2r.
\end{aligned}
\end{equation}
Analogously, the master equation for the blue individuals yields a similar
expression. These two PDEs in their expanded form cover approximately one page
when printed. While this is still a bit unhandy for a human being, it is not
at all a challenge for a computer. However, when we turn to more involved
examples, it is worthwhile to spend a few thoughts on the implementation. As
demonstrated above, expanding the equation after having inserted the Taylor
series, results in the large expression~\eqref{expand}, but most of its terms
are deleted by the polynomial reduction, giving~\eqref{polred}. In the present
example this is not a big deal, but in other cases this large intermediate
expression turns out to be the bottleneck. It is then advantageous to
systematically perform expansion--reduction steps on subexpressions; more
precisely, to follow a bottom-up approach, starting at the leaves of the
expression tree.

\subsection{Integration}

A common problem in symbolic computation is to bring the output of a
computation into a form that is useful for a human being. The computing power
that nowadays computers have allows to produce gigantic symbolic expressions
without much effort. It can be much more difficult to extract the relevant
information from such an output. In this spirit, we want to process the
Taylor-expanded expressions such as~\eqref{polred} further, and rewrite them
in a conservative, more compact form.

One of the classical problems in symbolic computation is to determine the
antiderivative of a given function. The first complete algorithm for the class
of elementary functions was given by Risch~\cite{Risch69}, which was later
extended to more general classes of functions, see for
example~\cite{Bronstein1997}. Most of these algorithmic ideas found their ways
into current computer algebra systems.

In contrast to the classical integration problem, we shall consider cases that
are more general, namely in the following three aspects.  First, the given
function~$a(x)$ may not have an antiderivative in the prescribed class; in
this case, it is desirable to decompose $a(x)$ into an integrable part and
remainder, i.e.,
\begin{equation}\label{decomp}
  a = \pa_x I + R,
\end{equation}
where the remainder $R$ is ``as small as possible''.  Second, the expressions
we are dealing with involve unspecified functions, so that the input can be
interpreted as a differential polynomial~\cite{Ritt50,Kolchin1973}; for example, we
would like to write the expression $f\cdot\pa_x f$ as $\pa_x\bigl(\frac12f^2\bigr)$.
Third, our setting is multivariate, in the sense that we have several
unspecified functions and several variables with respect to which we
differentiate.

The first algorithmic approach to the problem of integrating expressions with
unspecified functions was proposed in~\cite{Campbell1988}, and independently
for differential polynomials in~\cite{Bilge1992}. This was generalized
recently to integro-differential
polynomials~\cite{RosenkranzRegensburger2008,RosenkranzRegensburgerTecBuchberger2012},
to differential fields~\cite{Raab2012,Raab2013}, and to fractions of
differential polynomials~\cite{BLRR13,BKLPPU}.

While current computer algebra systems are very good in computing the
antiderivative of an expression involving unspecified functions (provided
that it exists), the decomposition into an integrable part and remainder
is a more delicate task. For example, both Mathematica and Maple correctly
compute
\begin{multline*}
  \int \bigl(f^2(\pa_x^2g) - 2(\pa_xf)^2g - 2f(\pa_x^2f)g\bigr)\,\mathrm{d}x = \\
  f^2(\pa_xg) - 2f(\pa_xf)g.
\end{multline*}
In contrast, if a given expression cannot be written as the derivative of some
other expression, then it is not at all straightforward to obtain a
decomposition of the form~\eqref{decomp}, using the standard integration
commands provided by the computer algebra system. As an example, consider the
decomposition
\[
  f\cdot\pa_xf+f = \pa_x\bigl(\tfrac12f^2\bigr) + f.
\]

We are now going to recall the main algorithmic ideas how to compute a
decomposition of the form~\eqref{decomp}. Let us first consider a
polynomial expression~$E$ in a single unspecified function~$f$ and its derivatives
$\pa_xf,\pa_x^2f,\dots$; let $n$ denote the order of the highest derivative
of~$f$ that appears in~$E$. If $E$ has an antiderivative, i.e., $E=\pa_xI$ for
some polynomial expression~$I$, then it is easy to see that $\pa_x^nf$ occurs linearly
in~$E$, i.e., $E$ is quasi-linear. Hence, if $\pa_x^nf$ does not occur
linearly, then the corresponding monomials are put into the remainder, as they
cannot be integrated. Now assume that $E$ is linear in $\pa_x^nf$. Let $m$ be
the highest power of $\pa_x^{n-1}f$ and denote by $u$ the coefficient of
$(\pa_x^{n-1}f)^m(\pa_x^nf)$ in~$E$, which is itself a polynomial in
$f,\pa_xf,\dots,\pa_x^{n-2}f$. Then integration by parts yields
\begin{equation}\label{intparts}
\begin{aligned}
  & u\cdot \bigl(\pa_x^{n-1}f\bigr){}^m(\pa_x^nf) = \\
  & \qquad \pa_x\Bigl(\frac{u}{m+1}\bigl(\pa_x^{n-1}f\bigr){}^{m+1}\Bigr)
    - \frac{\pa_xu}{m+1}\bigl(\pa_x^{n-1}f\bigr){}^{m+1}.
\end{aligned}
\end{equation}
Hence the first term on the right-hand side of~\eqref{intparts} goes into the
integrable part, while the second term is used to replace 
$u\cdot\bigl(\pa_x^{n-1}f\bigr){}^m(\pa_x^nf)$ in~$E$.
After performing this step repeatedly (at most $m$ times),
$E$ involves only derivatives of $f$ up to order~$n-1$. This shows
that the algorithm terminates.

We have seen that in the case of a single unspecified function, there is a
canonical choice which term to integrate in each step of the algorithm. In
contrast when several unspecified functions are involved, the situation is
less clear, as the following example shows:
\begin{equation}\label{intfg}
\begin{aligned}
  (\pa_xf)(\pa_xg)
  &= \pa_x\bigl(f(\pa_xg)\bigr) - f(\pa_x^2g) \\
  &= \pa_x\bigl((\pa_xf)g\bigr) - (\pa_x^2f)g.
\end{aligned}
\end{equation}
Hence one has to specify an order in which the terms are processed, and which
at the same time doesn't lead to infinite loops. The same kinds of problems
are faced when the unspecified functions depend on several variables. The
following example demonstrates the ambiguity of the decomposition in the
case of a single unspecified function~$f(x,y)$:
\begin{align*}
  & (\pa_xf)(\pa_yf)+\pa_xf+\pa_yf \\
  &\qquad = \pa_x\bigl(f\cdot \pa_yf+f\bigr) + \pa_y\bigl(f\bigr) - f\cdot\pa_x\pa_yf \\
  &\qquad = \pa_x\bigl(f\bigr) + \pa_y\bigl(f\cdot\pa_xf+f\bigr) - f\cdot\pa_x\pa_yf.
\end{align*}

In our application we have to deal with several unspecified
functions~$f_1,\dotsc,f_k$ in several variables, say $x,y,\dotsc,z$. So the
question is in which order we should treat the terms of the input expression
to obtain the desired result. One natural choice is to consider the variables
in a fixed order as the main loop of the algorithm. This means that we first
decompose the input with respect to the first variable, say $\pa_xI+R$; then
the remainder~$R$ is decomposed with respect to the next variable, and so on,
yielding a result of the form
\[
  \pa_xI_x + \pa_yI_y + \dots + \pa_zI_z + R.
\]
Additionally, one can also decompose~$I$ further, yielding a nested
decomposition of the following form (we show only the case of a single
variable):
\[
  \pa_x(\pa_x(\cdots(\pa_x(I)+R_d)+\cdots+R_2)+R_1)+R_0.
\]
In our description of the algorithm we use
the parameter~$d$ to specify the desired maximal integration depth of the
output expression.

For each integration variable, say~$x$, we proceed as follows: we determine
the highest derivative with respect to~$x$ that occurs in the input, no matter
which function is involved. We say that the highest $x$-derivative is of
order~$n$ if $\pa_x^nf_i$ occurs for some $1\leq i\leq k$, but there is no
index~$i$ such that $\pa_x^{n+m}f_i$ for some $m\geq 1$ occurs. Then for each $f_i$, $1\leq i\leq
k$, (in the order as specified by the user) the terms involving $\pa_x^nf_i$
are treated. Note that in this step derivatives with order $n+1$ can be
produced, as can be seen in~\eqref{intfg}. In order to avoid that the
algorithm runs into an infinite loop, we keep these terms, and continue by
considering derivatives of order~$n-1$. This algorithm is described in detail
in the following pseudo-code:
\medskip
\vspace{12pt}
\hrule
\noindent
\textbf{Algorithm} $\pai$\\
\textbf{Input:}
\begin{minipage}[t]{0.8\columnwidth}
  $E$: differential polynomial expression \\
  $f_1,\dotsc,f_k$: unspecified functions \\
  $x,y,\dotsc,z$: integration variables \\
  $d$: depth
\end{minipage}
\vspace{4pt}\hrule\vspace{4pt}
\begin{algorithmic}[1]
\setlength{\baselineskip}{13pt}
\If{$E=0$ \textbf{or} $\{x,\dotsc,z\}=\emptyset$ \textbf{or} $d=0$}
  \State \Return $E$
\EndIf
\State $R \gets 0$
\State $I \gets 0$
\State $n \gets$ highest $x$-derivative that appears in~$E$ for some $f_i$
\If{$n=0$}
  \State \Return $\pai\bigl(E, (f_1,\dotsc,f_k), (y,\dotsc,z), d\bigr)\!\!$
\EndIf
\newpage
\For{$i=n,n-1,\dotsc,1$}
  \For{$j=1,\dotsc,k$}
    \State $m \gets \expo\bigl(E,\pa_x^if_j\bigr)$
    \While{$m\geq2$}
      \State $g \gets \coef\bigl(E, \bigl(\pa_x^if_j\bigr){}^m\bigr)$
      \State $R \gets R+g\cdot \bigl(\pa_x^if_j\bigr){}^m$
      \State $E \gets E-g\cdot \bigl(\pa_x^if_j\bigr){}^m$
      \State $m \gets \expo\bigl(E,\pa_x^if_j\bigr)$
    \EndWhile
    \State $g \gets \coef\bigl(E,\pa_x^if_j\bigr)$
    \While{{$g\neq0$}}
      \State $m \gets \expo\bigl(g,\pa_x^{i-1}f_j\bigr)$
      \State $I \gets I+\frac{1}{m+1}(\pa_x^{i-1}f_j)g$
      \State $E \gets E-\frac{1}{m+1}\bigl((\pa_x^if_j)g+(\pa_x^{i-1}f_j)(\pa_xg)\bigr)$
      \State $g \gets \coef\bigl(E,\pa_x^if_j\bigr)$
    \EndWhile
  \EndFor
\EndFor
\State $R \gets R+E$
\State $I \gets \pai\bigl(I,(f_1,\dotsc,f_k),(x,\dotsc,z),d-1\bigr)$
\State $R \gets \pai\bigl(R,(f_1,\dotsc,f_k),(y,\dotsc,z),d\bigr)$
\State \Return $\pa_x(I)+R$
\end{algorithmic}
\hrule

When we apply our Mathematica implementation of algorithm $\pai$ to the large
expression~\eqref{polred} we obtain
\begin{align*}
\partial_t(r) &=
  \partial_x\bigl(r (b+r-1) (\alpha  r+1)\bigr) \\
  &\quad - (\gamma_1-\gamma_2) \partial_y\bigl(b r (b+r-1)\bigr) \\
  &\quad + h \Bigl(\tfrac12 \partial_x\bigl(\partial_x\bigl(r (\alpha  b r-b+\alpha  r^2-\alpha  r+1)\bigr)
    + 2 r \partial_xb\bigr) \\
  &\hspace{26pt} -(\gamma_1-\gamma_2) \partial_y\bigl(r (b+r-1) \partial_xb\bigr) \\
  &\hspace{26pt} +\gamma_0 \partial_y\bigl(2 r \partial_yb-\partial_y((b-1) r)\bigr) \\
  &\hspace{26pt} +\tfrac12 (\gamma_1+\gamma_2) \partial_y\bigl(r (2 b-r) \partial_yb-\partial_y((b-1) b r)\bigr)\Bigr)
\end{align*}
which basically agrees with the manually derived Equation~\eqref{e:pdered};
recall that $\rho=b+r$.  Comparing the two expressions reveals that
in~\eqref{e:pdered} some remainders are not minimal according to algorithm
$\pai$, but the overall expression is a bit more compact as it involves only
factored polynomials inside the derivatives.

\section{Numerical illustration for the mean-field model}

Finally we would like to illustrate the behavior of solutions
to~\eqref{e:pdesystem} with numerical simulations. We consider the
system~\eqref{e:pdesystem} on $\Omega\times (0,T)$, where in our computational
examples $\Omega=[-L_x,L_x]\times[-L_y,L_y] \subseteq \mathbb{R}^2$ with $L_y
\ll L_x$ corresponds to a corridor.  As individuals cannot penetrate the
walls, we set no flux boundary conditions on the top and bottom. At the
entrance and exit of the corridor, i.e at $x=\pm L_x$, we assume periodic
boundary conditions. For all numerical simulations we used the COMSOL
Multiphysics Package with quadratic finite elements. We set
$\Omega=[0,1]\times[0,0.1]$, choose a mesh of $608$ triangular elements and a
BDF method with maximum time step $0.1$ to solve the discretized system. The
first example models system \eqref{e:pdesystem} in the case where we have no
cohesion and no preference for stepping to one side, i.e. $\alpha=0$ and
$\gamma:=\gamma_1=\gamma_2$. In the second example we consider system
\eqref{e:pdesystem} with the special scaling
$\gamma_1-\gamma_2=\mathcal{O}(h)$ including cohesion and aversion. For this
particular scaling the first-order hyperbolic terms in $y$-direction are of
the same order as the diffusion in this direction while the mixed derivative
terms, i.e. the terms which involve derivatives with respect to $x$ and~$y$,
are of order $\mathcal{O}(h^2)$ and can be neglected.  In both simulations we
start with a perturbed configuration of the steady state, i.e. constant
densities for $r$ and~$b$, and study if the densities return to the constant
steady state or to another more complex stationary configuration.

\subsubsection{Example 1}
\begin{figure}[!ht]
\centering
\subfigure[Initial distribution of reds]{\includegraphics[width=63mm]{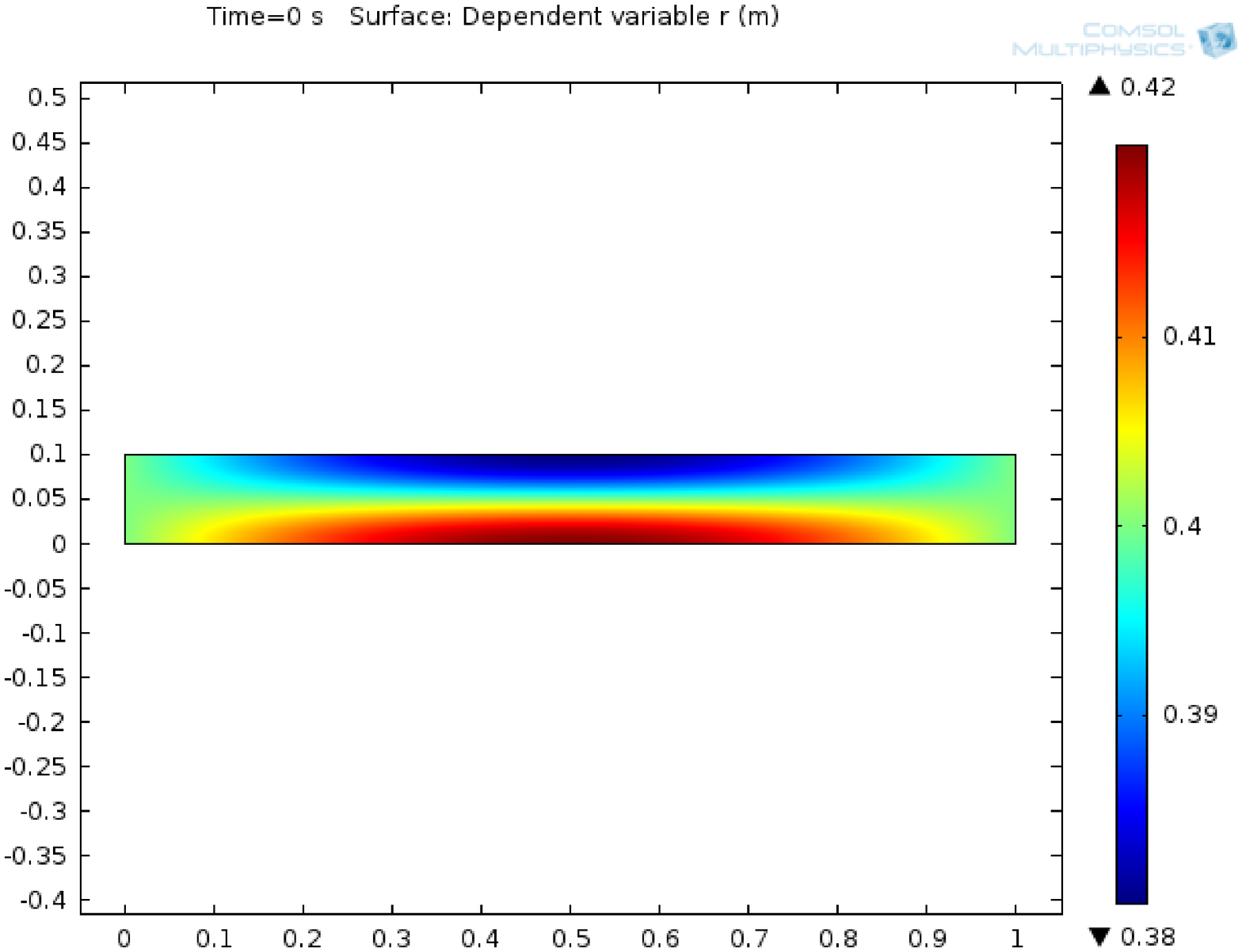}}
\subfigure[Distribution of reds at time $T=5$]{\includegraphics[width=63mm]{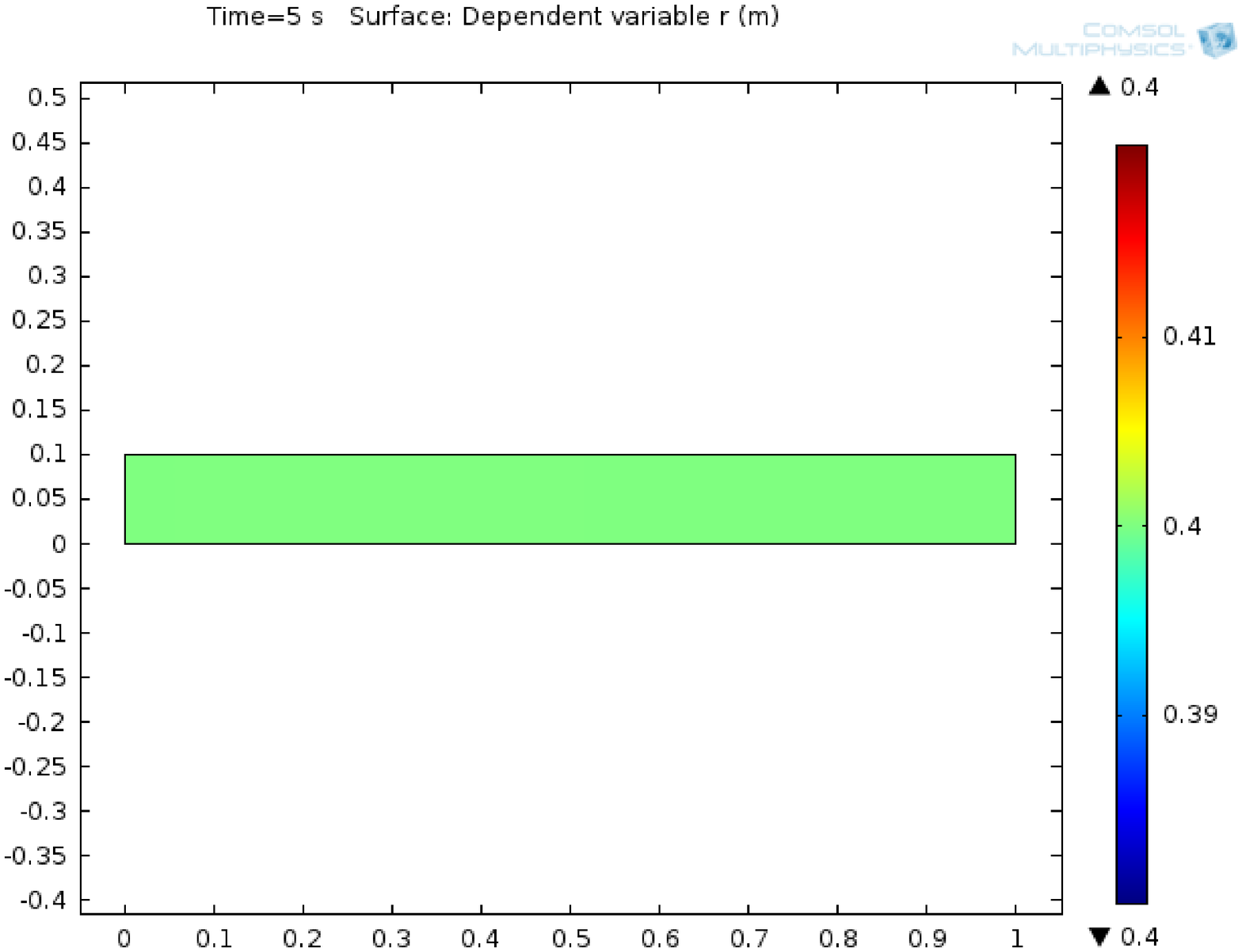}}
\caption{Solution to system~\eqref{e:pdesystem} with no cohesion and aversion.}\label{fig_ex1}
\end{figure}

Let $\gamma_0=0.1$, $\gamma=0.2$ and $h=0.3$. As initial values we choose small perturbations of an equilibrium state, i.e.
\begin{align}\label{perturb}
\begin{aligned}
r_0(x,y)&=0.4+0.02\sin(\pi x)\cos\left(\frac{\pi y}{0.1}\right)\\
b_0(x,y)&=0.4-0.02\sin(\pi x)\cos\left(\frac{\pi y}{0.1}\right).
\end{aligned}
\end{align}
The initial value $r_0$ and the solution $r_T$ to the system~\eqref{e:pdesystem}
at time $T=5$ is visualized in Figure~\ref{fig_ex1}. The corresponding density
of blue individuals show the same behavior, i.e. the densities return to the
constant equilibrium solution.
In this example we do not observe the formation of directional lanes as the
solutions return to the equilibrium state quickly.
\subsubsection{Example 2}
\begin{figure}[!ht]
\centering
\subfigure[Density of reds at time $T=5$]{\includegraphics[width=63mm]{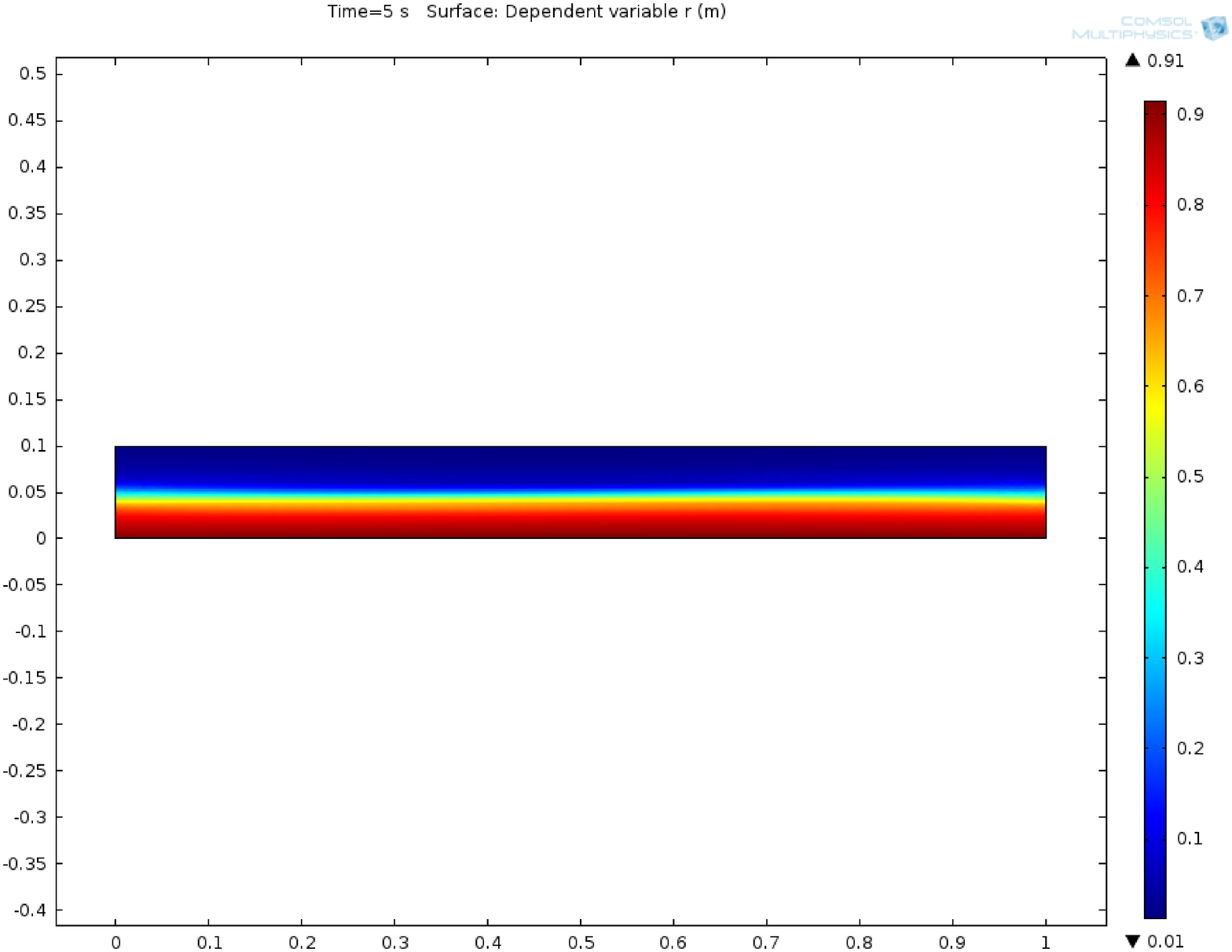}}
\subfigure[Density of blues at time $T=5$]{\includegraphics[width=63mm]{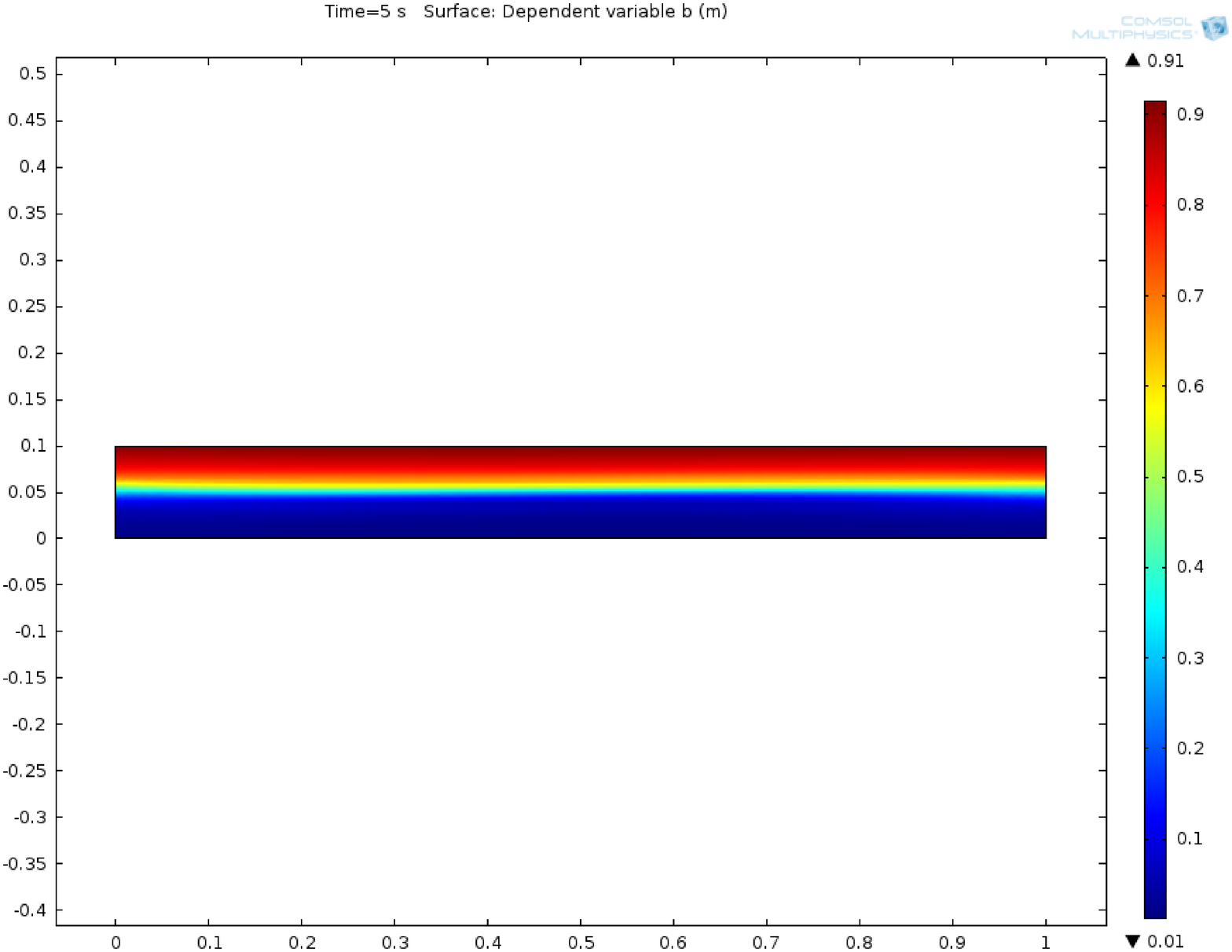}}
\caption{Solution to system~\eqref{e:pdesystem} with cohesion and aversion.}\label{fig_ex2}
\end{figure}

Let $\gamma_0=0.001$, $\gamma_1=0.5$, $\gamma_2=0.4$, $\alpha=0.2$ and
$h=0.1$, i.e. $\gamma_1-\gamma_2=\mathcal{O}(h)$. As initial values we choose
again~\eqref{perturb}, i.e. small perturbations of an equilibrium
state. Figure~\ref{fig_ex2} shows the solution $r_T$ and $b_T$ to
system~\eqref{e:pdesystem} at time $T=5$.
In this example we observe lane formation. Since $\gamma_1>\gamma_2$
individuals have a tendency to step to the right. This tendency can also be
observed in the formation of the directional lanes. The red individuals
concentrate on the bottom of the corridor, whereas the blue individuals move
to the top.

\section{Conclusion and outlook to further applications}
In this paper, we have presented a symbolic approach to derive the mean-field PDEs from lattice-based models. We demonstrated the methods in terms of one example in pedestrian dynamics, but the algorithm may also be applied to other
examples. In~\cite{FLS2010} different motility mechanisms on regular lattices
are introduced, which result in nonlinear diffusion equations with different
diffusivities. The authors considered various motilities based on attraction or
repulsion, i.e. the transition rate to move away from a neighbouring
individual increases or decreases respectively. For example, in a minimal model
the transition rate is given by
\begin{align*}
\mathcal{T}^{i \rightarrow i+1} = (1-c_{i+1}) (1 - \alpha c_{i-1}), 
\end{align*}
where $c_i$ again denotes the probability that the lattice site $x_i$ is
occupied. Hence the transition to move from $x_i$ to $x_{i+1}$ is reduced if
the neighbouring site $x_{i-1}$ is occupied.  This phenomenon is known as
adhesion and results in a nonlinear diffusion model for the cell density
$c = c(x,t)$ of the form
\begin{align*}
\partial_t c  = \partial_x (D(c) \pa_xc),
\end{align*} 
with a diffusivity of the form $D(c) = 3 \alpha (c - \frac{2}{3})^2 + 1 -
\frac{4}{3}\alpha$, see~\cite{anguige2009one}. Again in~\cite{FLS2010} several
other transition rates were proposed, which lead to different nonlinear
diffusitivies, see Table~1 there. Using our implementation, the entries in
that table can be generated automatically. For example, we have tried one of
their most complicated models~\cite[Equation~(13)]{FLS2010}, which combines
contact-forming or contact-breaking interactions with contact-maintaining
interactions. Our implementation correctly derives the diffusivity given
in~\cite[Equation~(14)]{FLS2010}, where we have chosen the two-dimensional
square lattice with Moore interacting neighborhoods.

\section*{Acknowledgment}
CK was supported by the Austrian Science Fund (FWF): W1214.
HR and MTW acknowledge support by the Austrian Academy of Sciences \"OAW via the New Frontiers project NST-0001.
GR was supported by the Austrian Science Fund (FWF): P27229.


\bibliographystyle{IEEEtran}
\bibliography{IEEEabrv,synasc}

\end{document}